\documentstyle[12pt,aasms4,psfig]{article}

\newcommand{\hetgs}{{\it HETG} }

\newcommand{\chandra}{{\it Chandra} }

\newcommand{\fekfiga}{Figure 1a }

\newcommand{\fekfig}{Figure 1 }
\newcommand{\figcont}{Figure 2 }
\newcommand{\figblr}{Figure 3 }
\newcommand{\etal}{{\it et al.} }
\newcommand{\asca}{{\it ASCA} }

\newcommand{\fekalfa}{${\rm Fe} \rm \ K\alpha$ }

\newcommand{\mcg}{MCG~$-$6$-$30$-$15 }

\newcommand{\psrc}{NGC~5548 }

\begin{document}

\title{
PHYSICAL DIAGNOSTICS FROM A NARROW  \fekalfa EMISSION LINE 
DETECTED BY CHANDRA IN THE SEYFERT 1 GALAXY \psrc}

\author{T. Yaqoob\altaffilmark{1,2},
I. M. George, \altaffilmark{1,2},
K. Nandra\altaffilmark{1,3},
T. J. Turner\altaffilmark{1,2},
P. J. Serlemitsos\altaffilmark{1}
R.~F.~Mushotzky\altaffilmark{1}}

{\it Accepted for Publication in the Astrophysical Journal (submitted July 5, 2000)}

\altaffiltext{1}{Laboratory for High Energy Astrophysics, 
NASA/Goddard Space Flight Center, Greenbelt, MD 20771, USA.}
\altaffiltext{2}{Joint Center for Astrophysics, Physics Department,
University of Maryland, 
Baltimore County, Baltimore,
1000 Hilltop Circle, MD21250.}
\altaffiltext{3}{Universities Space Research Association}

\begin{abstract}
We report the detection 
of a narrow  \fekalfa emission line in the Seyfert~1 galaxy NGC 5548 with the
\chandra High-Energy Transmission Grating. 
In the galaxy frame we measure a  center energy 
of $6.402 ^{+0.027}_{-0.025}$ keV, a
FWHM of $4515 ^{+3525}_{-2645}$ km/s,
an intensity of $3.6 ^{+1.7}_{-1.5} \times 10^{-5} \ \rm 
photons \ cm^{-2} \ s^{-1}$, with an
equivalent width of $133 ^{+62}_{-54}$ eV (errors are 90\% confidence
for one parameter).
The line is only marginally resolved at the 90\% confidence
level.
The line energy is consistent with an origin in cold,
neutral matter, but ionization states up to $\sim \rm Fe~XVIII$ are
not ruled out.
We cannot constrain the detailed dynamics but assuming
Keplerian motion, the velocity width is consistent 
with the 
line being produced in the outer optical/UV broad-line region (BLR) 
at about a light-month 
from the central X-ray source. 
We cannot rule out some contribution to the 
narrow \fekalfa line from a putative, parsec-scale obscuring torus
which is postulated to be a key component of AGN unification models. 
The continuum intensity during the
\chandra observation was a factor $\sim 2$ less than typical 
historical levels.
If the X-ray continuum was at least a
factor of 2 higher in the recent past before the \chandra
observation and the narrow \fekalfa intensity had not yet responded to
such a change 
then the predicted line intensity and equivalent width
for an origin in the BLR
is within the 90\% measurement errors. 
Anisotropic X-ray continuum illumination of the BLR and/or
additional line emission from a torus structure would
improve the agreement with observation.
Two out of three archival \asca data sets  
are consistent with the narrow line being present with the same 
intensity as in the \chandra observation. 
However, there is strong evidence that the narrow-line intensity varied
and was unusually low during one of the \asca campaigns.
In general, inclusion of the narrow line
to model the overall broad  \fekalfa line profile in terms of a rotating
disk plus black-hole model
can have a non-negligible effect on the disk-line intensity and variability
properties.
Variability of the
broad disk line in NGC 5548 is difficult to reconcile 
with the expectations of the simple disk model,
even when the
narrow-line component is accounted for.
It will be important to ascertain the
importance of a similar non-disk \fekalfa line in other Seyfert~1
galaxies.
Future monitoring of the narrow \fekalfa component with large collecting
area and high spectral resolution will enable 
reverberation mapping of the BLR region complementary to similar
studies using the optical/UV lines and therefore provide independent 
constraints on the black-hole mass.

\end{abstract}

\section{INTRODUCTION}

The {\it Advanced Satellite for Cosmology and Astrophysics}
(\asca -- Tanaka \etal 1994) found that the  \fekalfa 
fluorescent emission line in Seyfert~1 galaxies is 
often very broad 
and is generally interpreted as the result of an origin in
matter  
in an 
accretion disk rotating around a central black hole
(see Fabian \etal 2000 and references therein).
The line profile is sculpted by characteristic
gravitational and Doppler energy shifts.
Currently, study of the  \fekalfa emission line is
the only way to probe
matter within a few to tens of gravitational radii
of a black hole.
There is an important caveat with respect 
to studying the relativistic  \fekalfa line
profiles with {\it ASCA}. The energy resolution of the CCD detectors
is only $E/ \Delta E \sim 40$ (or FWHM $\sim 160$ eV, or $\sim 7500$ km/s)
in the  \fekalfa band (6--7 keV).
Thus, even for the highest signal-to-noise
data, one could not {\it unambiguously}
separate that part of the line
which originates in the accretion disk, from any other component
originating in matter beyond the accretion disk, further away 
(e.g. see Weaver and Reynolds 1998; Done, Madejski, and Zycki 2000).
Such line components will be much narrower than the disk line, due to
the lower Keplerian velocities and small gravitational shifts.
Indeed, such a narrow  \fekalfa emission line
from the putative obscuring torus is thought to be the key for unifying type 1
and type 2 AGN, and is 
expected from Seyfert~1 galaxies (e.g. see
Krolik, Madau, and Zycki 1994; Ghisellini, Haardt, and Matt 1994).
The torus
is extended on the parsec-scale and would lead to a FWHM for the
narrow  \fekalfa line from this region of
$\sim 760 \sqrt{(M_{8}/r_{\rm pc})}$ km/s 
(assuming a virial relation and an r.m.s. velocity
dispersion of $\sqrt{3} V_{\rm FWHM}/2$; see e.g. Netzer 1990).
On the same scale,
narrow-line emission may also originate
in the extended warm scattering region which is also a feature of
unified AGN models (e.g. see Krolik and Kallman 1987).
A broader narrow-line contribution
(1000s of km/s) might also be expected from the
optical/UV broad-line region (BLR).

As we shall demonstrate in this paper,
accurate  measurements of the narrow  \fekalfa line energy, width and
intensity can provide a suite of powerful physical diagnostics of the 
central engine in AGN. 
Up to now, 
the expected narrow  \fekalfa line has never been unambiguously detected
in a Seyfert~1 due to limited instrumental energy resolution and
sensitivity. However, strong evidence  for its existence 
and physical parameters of the putative line  
have been deduced 
for several sources (e.g. Weaver \etal 1993;
Yaqoob \& Weaver 1996; Yaqoob \etal 1995,1996;
Sulentic \etal 1998; Weaver \& Reynolds 1998; Done \etal 2000).
The narrow line appears to be most prominent in NELG such as NGC 2992
(Weaver \etal 1996), which may sometimes have an underlying broad component
(e.g. MCG $-$5-23-16, Weaver \etal 1997).
The above evidence for the narrow \fekalfa line in Seyfert~1 galaxies
has been tempered by the case of
\mcg
in which, during a flare,
the entire broad  \fekalfa line was highly redshifted, 
leaving no clear
emission peak around 6.4 keV (Iwasawa \etal 1999). This has
led to some debate about the
existence of a narrow component. 

In this paper we
report the detection of a narrow \fekalfa line in the Seyfert~1
galaxy NGC 5548 from a \chandra observation.
In \S \ref{chdata} we present the data and describe the analysis
procedures. In \S \ref{detected} we
show that the \fekalfa line in NGC 5548 cannot be explained
by emission from a relativistic disk alone and
present the basic measurements
of the narrow \fekalfa line parameters. In \S \ref{broad} we discuss all the 
different possible mechanisms contributing to the width of the
narrow line and give our best estimate of its velocity  width.
We discuss the physical constraints imposed by the narrow line in
\S \ref{diagnos}, and in \S \ref{ascadata} we investigate the
effect of the narrow line on the parameters derived for the
relativistic broad line using archival \asca data.
In \S \ref{variability} we discuss some implications of the presence
of a non-disk \fekalfa component on the interpretation of
variability studies of the line and continuum.
Finally, we present our conclusions in \S \ref{concl}.

\section{THE CHANDRA DATA}
\label{chdata}

NGC 5548 was observed with \chandra starting on 2000 February
5, UT 15:38 for a total duration of 83 ks (continuous). 
The instrument used in the focal plane of the
High Resolution Mirror Assembly (HRMA) was the High-Energy
Transmission Grating (or \hetgs -- 
Markert, \etal 1995).
The \hetgs consists of two grating assemblies, a High-Energy
Grating (HEG) and a Medium-Energy Grating (MEG). The HEG 
has a better efficiency than the MEG above $\sim 5$ keV.
The spectral resolution of the HEG is roughly two times better
than the MEG ($\Delta \lambda = 0.012 \AA$ and $0.023 \AA$
for the HEG and MEG respectively). $\Delta \lambda$ is 
roughly constant
over the instrument bandpass (HEG: $\sim 0.8-9$ keV , MEG:
$\sim 0.5-9$ keV). Events dispersed by the gratings are collected
by a CCD array and can be assigned  an energy based on the
position along the dispersion axis. Since the CCDs have
intrinsic energy resolution, background events can be rejected
with a high efficiency, and different spectral orders can be easily
discriminated.

Our \chandra observation of NGC 5548 yielded a net exposure
time of 82,198.4 s. After accounting for a deadtime factor of
0.987337, the effective exposure time is 81157.6~s
For our analysis we used data from events 
reprocessed using the latest software and calibration, as of
3 July 2000
\footnote{Details of the \chandra pipeline reprocessing and
data analysis specifics can be found at
http://asc.harvard.edu/ciao/threads/}.
Genuine photon events collected by the CCDs fall into  
specific pixel patterns classified by their {\it grade} and 
we retained only grades 0, 2, 3, 4, and 6; rejected grades correspond
to non-X-ray events.
We examined the
radial profile of the zeroth-order image and found a single Gaussian
to be a poor representation, due to the effects of
pile-up broadening the profile  (the best-fitting width was
$\sigma = 0.52$ arcseconds). A double Gaussian model gave a good fit and
the best-fitting widths of two Gaussian were $0.48 \pm 0.01$ and 
$1.09 \pm 0.07$ arcseconds,
the ratio of the normalization of the narrow to broad Gaussian
being 27.4.
Since the zeroth order data suffer from pile-up 
we did not use these data for spectral analysis. 
For the grating data,
in the cross-dispersion direction we found that the counts distribution
as a function of cross-dispersion angle could be approximated
by Gaussians with widths in the range 0.47--0.58 arcseconds for the
four arms of first-order data 
($\pm 1$ orders for the HEG and MEG). 
We accumulated grating spectra from events along the
dispersion direction, and within  $\sim \pm 3.6$ arcseconds of the 
peak in the cross-dispersion
direction. 
In the first-order spectra we obtained total count rates
(summed over the $\pm 1$ orders)
of 0.19 cts/s and 0.43 cts/s for the HEG and MEG respectively.
The count rates in the higher order spectra are at most 11.3\% and
9.8\% of the first order count rates for HEG and MEG respectively
so in this paper we will restrict our analysis to the first order
spectra only.

We do not expect further refinements to the CCD gain 
to affect the energy scale of the grating data 
since the energy assignment of events depends
only on the position of the event on the CCD 
according to  the dispersion relation. The CCD energy scale may, however
affect the rejection of source events as
background events and vice versa. Since we
are most concerned with the  \fekalfa band and this
band will have the highest proportion of
background in the data that is relevant for this
paper, we checked how important background
is for our data in the 5.5--7 keV band. 
We collapsed the 5.5--7 keV first-order HEG data in the dispersion
direction and examined the count distribution as a function of
cross-dispersion angle, which is again approximately Gaussian. 
We  
estimated the background level by fitting a model consisting of
a Gaussian plus
a constant and found that the background is  
$\sim 1.5 \%$ of the total counts in the 5.5--7 keV band. The same procedure
for the MEG showed that the background is $\sim 2.2\%$ of the total 
counts in the 5.5--7 keV band.

We made effective area files (ARF, or {\it ancillary response file})
using {\sc ciao} v 1.1 
({\it Chandra Interactive Analysis of Observations}), 
which also takes account of the dithering of the
satellite. 
We only used  {\sc ciao} for the reprocessing of the event data
and to make the ancillary response files; the remaining data analysis
(except for spectral fitting)
was done using IDL routines written by ourselves. 
For the spectral fitting procedures described
below, we used spectral response matrices 
{\tt acisheg1D1999-07-22rmfN0002.fits} (HEG) and
{\tt acismeg1D1999-07-22rmfN0002.fits} (MEG). 
Note that
absolute flux-calibration may be uncertain by as much as
30\% at the time of writing (Marshall \etal 2000). However our principal results
(i.e. center energy, width and equivalent width of the \fekalfa line)
do not depend at all on the absolute flux calibration.

The spectral energy resolution is determined principally
by the telescope response function but 
could be degraded by grating period variation (only at very low energies),
incoherent dispersion ($\sim 1\%$ of first order events), incorrect
aspect solution, and defocussing. 
By examining the zeroth-order image from CCD frame-transfer events
(which do not suffer pile-up), Paerels \etal (2000) determined 
from a \hetgs observation of Cyg X-3 that
these effects do not compromise the spectral resolution within the
measurement uncertainties. 
We also examined the frame-transfer events (or `streak data') of
our NGC 5548 data
and found that the photon distribution in the dispersion
direction could be approximated well with a single Gaussian, resulting
in a best-fitting width of $\sigma=0.41 \pm 0.03$ arcseconds
(90\% confidence errors), consistent
with nominal expectations. This reassures us that there are no
adverse effects on the spectral resolution of the gratings.
The
absolute calibration of the wavelength scale is typically
$\sim \rm 2m\AA$ (e.g. Paerels \etal 2000), 
much smaller than the statistical uncertainty in the
 \fekalfa emission line energy that we are
concerned with in this paper.
 
\section{DETECTION OF A NARROW  \fekalfa LINE IN NGC 5548}
\label{detected}

We used XSPEC for spectral fitting
the 2--7  keV HEG and MEG spectra
(the signal-to-noise for both the HEG and MEG drops rapidly above
$\sim 7$ keV). The $-1$ and $+1$ orders were summed for the HEG and MEG.
The spectral bins had a width of $\sim 0.005 \AA$ which corresponds
to the $1\sigma$ width of the spectral resolution of the HEG. For
most of the 2--7 keV bandpass, in the HEG or MEG spectra,
there were less than 10 counts per bin
and at most 14 or 12 counts per bin (HEG and MEG respectively).
Since the data are in the Poisson regime we 
performed the spectral fitting and error analysis using the 
C-statistic. 
\fekfiga shows the HEG and MEG spectra fitted with a simple
power-law model, which describes well the overall spectra in the 2--7 keV
band. 
The consistency between the HEG and MEG is good over most of this
energy range, except that between $\sim 4-5$ keV the MEG shows
systematic deviations of the order of $\sim 20-30\%$.
\fekfiga also shows
that an  \fekalfa emission line is clearly detected
with high signal-to-noise in both the HEG and the MEG.
The insets in \fekfig show close-ups of the  \fekalfa region 
for the HEG data, and also the  \fekalfa region from an \asca observation
in July 1996.
The \asca \fekalfa line is broad, with no obvious
signature of a narrow core (but note that the \asca energy
resolution in the Fe-K region is $\sim 160$ eV FWHM, a factor of
$\sim 4$ worse than the HEG data in that band).

When we added a Gaussian 
to the power law
continuum to model the \fekalfa emission line
in the HEG plus MEG spectra, the C-statistic decreased by 25.6,
which corresponds to a detection significance of nearly $5\sigma$
for three degrees of freedom.
In this latter model,
there were six free parameters in total; two independent normalizations
for the continuum (one for the HEG and one for the MEG),
the power-law photon index ($\Gamma$), the line intensity, Gaussian width
and center energy. 
The results of this spectral fit are shown in Table 1, along with the
90\% confidence, one-parameter errors for the derived model parameters. 
Also shown in Table 1 for reference are the spectral
fitting results obtained when the same model is applied to the
HEG data only. These latter results are consistent with the joint
fit so hereafter we will refer only to the parameters derived 
from the combined
HEG and MEG fits.
We obtain a photon index for the power
law of $1.54 \pm 0.06$ and
a 2--10 keV flux of $2.4 \times 10^{-11} \rm \ ergs \ cm^{-2} \ s^{-1}$.
Historical 2--10 keV flux levels for NGC 5548 measured
by {\it EXOSAT}, {\it Ginga}, and {\it ASCA}, according to
the archival HEASARC databases and published results, span the
range $1.7-6.4 \ \times 10^{-11} \rm \ ergs \ cm^{-2} \ s^{-1}$,
with values in the upper third of this range being most common
(e.g. see Turner \& Pounds 1989; Nandra \& Pounds 1994; Chiang \etal 2000). 
Thus it appears that the 2--10 keV flux during the \chandra observation
was a factor of $\sim 2$ down from typical historical values.

Unless otherwise stated, we will refer all our
measurements for X-ray line parameters to the galaxy frame, using
a redshift of $0.01676$, obtained from the NASA/IPAC Extragalactic
Database (NED). Note that the redshift of NGC 5548 is uncertain by
the equivalent of $\sim 100 \rm \ km/s$, and the value $z=0.01676$ refers
to the value otained from optical emission-line measurements,
whilst $H\sc{i}$ measurements give $z=0.017175$. See, for example,
Mathur, Elvis, and Wilkes (1999) and references therein.
Thus, our quoted galaxy-frame \fekalfa 
line-energy measurements carry additional
uncertainties of the order of $\sim 2$ eV.
 
Table 1 shows that for the Gaussian fit
we obtain a
peak energy, intensity, and width of the 
narrow \fekalfa line detected by the HEG and MEG, of 
$6.402 ^{+0.027}_{-0.025}$ keV, 
$3.6^{+1.7}_{-1.5} \times 10^{-5} \rm \   photons \ cm^{-2}
\ s^{-1}$, and $4515^{+3525}_{-2645}$ km/s FWHM respectively.
The equivalent width (EW) of the line is $133^{+62}_{-54}$ eV. 
Statistical errors are $90\%$ confidence for one parameter for
all the above measurements. 
Yaqoob (1998) noted that, in the extreme
Poisson limit, the $C$-statistic can under-estimate the
statistical errors on spectral features which are comparable to
or narrower than 
the instrument resolution. Although our spectra are not in the
extreme Poisson limit, we performed simulations (as described in
Yaqoob, 1998) to check the reliability of the statistical
errors derived from spectral fitting.  
Thus, we made a grid of simulated spectra
in which the input \fekalfa line intensity was set at nine values in the
range 1.5--6.0 $\times 10^{-5} \rm \ photons \ cm^{-2} \ s^{-1}$ and the
input (Gaussian) line width was set at twelve values in the range 4--80 eV. The
input line energy was set at 6.402 keV in the galaxy frame (the
best-fit value from spectral fitting). 
Each grid
point contained 1000 simulated spectra. 
All other model parameters (e.g. exposure time)
were set at the values obtained from the \chandra data. 
The simulated spectra were fitted with the same power law plus
Gaussian model and we found the statistical errors on the derived
model parameters to be consistent with those obtained from spectral
fitting. 
Our simulations also show that
the HEG data contain a mean of only {\it 70 photons in the \fekalfa line
with a $1\sigma$ dispersion of $\pm 8$ photons}.
This is consistent with the mean number of line photons (72) we obtain
from the product of the effective area at the observed line center
energy, the exposure time and the best-fitting line intensity from
Table 1. A similar calculation shows that the expectation 
value of the number of line photons detected by the MEG is 23. 

We next addressed the question of whether the narrow \fekalfa line
detected by \chandra is resolved. 
Note that the spectral resolution of the HEG at the 
observed \fekalfa line energy corresponds
to $\sim 38$ eV FWHM, or $\sim 1800$ km/s FWHM.
Starting with the best-fitting
power law plus Gaussian model to the HEG and MEG data
we constructed confidence contours using the C-statistic for
the \fekalfa line intensity versus the line width. \figcont shows 
the 68\%, 90\% and 99\% joint confidence regions, from which it 
can be seen that the emission line is resolved at the 90\% level
but not at 99\%.

\subsection{RELATIVISTIC DISK LINE CONTRIBUTION}
\label{diskinaxaf}

We note that the \fekalfa emission line detected by \chandra cannot
be accounted for by a relativistic disk line alone. We fitted the
HEG plus MEG data with a power-law continuum and
a model of the broad \fekalfa line originating in an
accretion disk rotating around a Schwarzschild black hole (e.g. Fabian
\etal 1989). 
We fixed the disk power-law emissivity index at $-2.5$
(close to the mean of a sample of Seyfert~1 galaxies -- Nandra \etal 1997),
and the energy of the line in the disk rest-frame at 6.4 keV.
The inner and outer disk radii  were
fixed at $6_{r_{g}}$ and $1000 r_{g}$
respectively, where $r_{g}=GM/c^{2}$, and $6_{r_{g}}$ is the radius 
of the last marginally stable orbit around a black hole of mass $M$.
The data are not very sensitive to the parameters that were fixed.
The inclination angle ($\theta_{\rm obs}$) and the disk-line intensity
were allowed to float. Including two independent continuum normalizations
for the HEG and MEG, there are only a total of five free parameters,
one less than the power law plus Gaussian line model.
The best fit was much worse than the one in which the \fekalfa line
was modeled only by a Gaussian (see Table 1); 
specifically the C-statistic was
higher by 10.1 for the disk-line model, for one less 
degree of freedom. This means that the confidence level for the
Gaussian model being a better description of the line is greater than
99.8\%.
The best-fitting
disk inclination was $14.6^{+4.6}_{-4.0}$ degrees 
(90\% confidence for one parameter), which is 
inconsistent with the best-constrained inclination derived from
fitting \asca data (see \S \ref{ascadata}).
Essentially, the narrow \fekalfa line detected by \chandra is
too sharply peaked for even low-inclination disk-line models.

Next, we investigated the effect of including a broad, relativistic
\fekalfa line in addition to the narrow Gaussian model. This time,
in addition to the parameters which were fixed
in the disk-line only fits, we also
fixed the disk inclination angle at the
best-constrained value from the \asca
data, of $31^{\circ}$ (see \S \ref{ascadata}).
All the Gaussian parameters were still free so there is only one
additional degree of freedom in this composite, dual-line model.
Spectral fitting the HEG and MEG data with this model
reduces the C-statistic by only 0.1 relative to the
power law plus Gaussian model. We note here that the \chandra data
are insensitive to the disk inclination angle, with the C-statistic changing
by less than 0.6 when the inclination angle is varied over
the entire range of $0^{\circ}$ to $90^{\circ}$. 
Since the relativistic line is so broad, its flux is spread out over
a wide energy range and we found that it has a negligible effect on the
derived parameters of the narrow \fekalfa line. In particular, the
intensity of the narrow line decreased by $\sim 3\%$ and the width increased
by $\sim 4\%$ when the
disk line was included. The best-fitting intensity and equivalent
width of the relativistic disk line was $0.4^{+7.2}_{-0.4}
\times 10^{-5} \rm \ photons \ cm^{-2} \ s^{-1}$ and
$13^{+227}_{-13}$ eV respectively (errors are
90\% confidence for three parameters). Thus, the broad component of the
\fekalfa line is not detected due to the small effective area of the
\chandra gratings but the statistical upper limits on its intensity
are consistent with historical \asca measurements (Chiang \etal 2000,
and \S \ref{ascadata}).

\section{BROADENING MECHANISMS FOR THE NARROW \fekalfa LINE}
\label{broad}

The intrinsic 
width of the \fekalfa line, which results from a combination
of the natural width and level width, is less than 1 eV 
(Bambynek 1972) so for our purpose it is negligible.
The thermal width of the narrow \fekalfa line is $\sim 1.7\sqrt{(
T/10^{4} K)}$ km/s so this too is negligible.
We can also neglect gravitational redshifts (these 
are of order $\Delta E/E \sim r_{g}/r$ for $r/r_{g}  \ll 1$,
where $r_{g} \equiv GM/c^{2}$)
which will be less than 
1 eV at a distance greater than four light days from a black hole
with a mass  of $10^{8} M_{\odot}$). In \S \ref{diagnos} and \figblr
we show that the narrow \fekalfa line originates at $\sim 1$ light
month from the central X-ray source.

The \fekalfa line actually consists of two lines, K$\alpha_{1}$
at 6.404 keV and K$\alpha_{2}$ at 6.391 keV, with a branching ratio
of 2:1. Therefore 
we investigated how this affects our results as follows.
Using the 2--7 keV HEG and MEG data 
we fitted a model consisting of  a power-law continuum plus two 
Gaussians, with the center energies fixed at the above energies,
and the intensity of the
Gaussian at higher energy was constrained to
have twice the intensity of the other Gaussian.
The widths of the two Gaussians were constrained to vary together.
We obtained best-fitting values for the total line intensity
(the sum of the two Gaussian intensities) and the width which were
the same as those obtained from a single Gaussian model (Table 1), and also
obtained similar statistical errors. In particular we 
obtained a width of the line of $4515 ^{+3630}_{-2530}$ km/s
FWHM (compare with the single-Gaussian measurement in Table 1).

If the narrow \fekalfa line is formed in Compton-thick
material such as the putative obscuring
torus, Compton down-scattering will also increase
the apparent width of the line. However the Compton shoulder
on the red side of the line peak due to the first scattering
is already an order of magnitude down in intensity relative to the
unscattered component (e.g. Matt, Brandt, and Fabian 1996; 
Iwasawa, Fabian, and Matt, 
1997). The present data are not sensitive to this level of
structure. 

Another uncertainty arises from a possible contribution to the \fekalfa
line from higher ionization states such as Fe II, which is expected if
some of the line is coming from the putative obscuring torus. However,
laboratory measurements of the energies of these higher ionization lines is
uncertain and the HEG does not have sufficient energy resolution to
place any tighter constraints other than the fact that the 
90\% confidence errors
on the line center energy rule out a dominant contribution from
ionization states higher than Fe~XVIII (e.g. Makishima 1986).  
Thus, until better data are available
we attribute the broadening of  
the narrow \fekalfa line in the \chandra data for NGC 5548 
to the bulk velocity of the emitting gas but do not rule out some
contribution 
from emission components with velocity widths less than our
formal lower limits imply. 

\section{PHYSICAL DIAGNOSTICS FROM THE NARROW  \fekalfa LINE}
\label{diagnos}

The narrow  \fekalfa line discovered in 
NGC 5548 immediately enables some robust deductions to be made.
We showed in \S \ref{diskinaxaf} that line is too narrow to be the core of
a relativistic disk line (e.g. see Fabian \etal 1989, 2000)
in which the line-emissivity falls off
with increasing distance from the continuum source.
In the following sections we use
the measurements of the peak energy, width, intensity and
equivalent width of the narrow \fekalfa line
to place physical constraints on the line-emission region. 

\subsection{PEAK ENERGY OF THE LINE}
\label{peak}

The peak energy of the line is $\sim 6.40 \pm 0.03$ keV, 
consistent with the rest-energy for
cold, neutral iron. The upper limit on the energy
corresponds to an ionization state of Fe~XVIII and this is
consistent with the ionization parameter for the optical/UV 
BLR deduced by Goad and Koratkar (1998) of $\log{U} \sim -0.6-0.0$
from detailed modeling.
As mentioned in \S \ref{broad},
gravitational redshifts are not large enough 
to significantly shift back the 
line energy from ionized Fe to lower energies (the distance of
the line-emitting region we deduce below produces energy
shifts less than 1 eV). 
Therefore a dominant emission
component from a warm scattering zone,
consisting of He-like or H-like Fe, (a component of
type 1 and type 2 AGN unification models, e.g. Krolik \& Kallman, 
1987) appears to be ruled out,
at least in NGC 5548. Models in which the {\it broad}  \fekalfa line
is produced and broadened by Compton scattering
in an optically thick medium 
(e.g. Misra and Kembhavi 1998; Mouchet \etal 2000) are likewise ruled out since
the unscattered peak of the  \fekalfa line would be at $\sim 6.67$ keV.

We can also rule out a scenario in which the narrow \fekalfa line
is not seen directly but from reflection in optically thin cold gas
out of the line-of-sight. Line photons of energy $E_{0}$ will be
Compton down-scattered to a mean energy of
$E_{0}/[1 + E_{0}/m_{e}c^{2}]$.
For $E_{0} = 6.4$ keV this is $\sim 80$ eV lower than the
peak energy we measure from the HEG data for NGC 5548. Of course
if the scattering medium is hot, Fe XX could produce a line which is
scattered back down to the observed energy. However, this requires
a temperature of the order of 1 keV, and this would give rise to
a Compton-broadened line with a FWHM $\sim 670$ eV,
an order of magnitude larger than observed
(see Pozdnyakopv, Sobol and Sunyaev, 1983). In fact, to obtain a
line width consistent with the observed value, the gas electron temperature   
would have to be less than $1.4 \times 10^{5}$ K.

\subsection{LINE WIDTH}
\label{lwidth}

From a detailed study of the velocity widths and line/continuum
time lags for 
optical/UV emission lines in NGC 5548, Peterson \& Wandel (1999)
found strong evidence for Keplerian motion of the line-emitting
matter and estimated the central mass to be $6.8 \times 10^{7}
\ M_{\odot}$.
As Peterson \& Wandel (1999) caution, this mass estimate is 
subject to uncertainties in the geometry and kinematics of the BLR,
and optical/UV line reprocessing physics. The actual mass may be as
small as $\sim 5 \times 10^{6} \ M_{\odot}$ and as high as
$\sim  10^{8} \ M_{\odot}$. A summary of
estimates of the central mass in NGC 5548
using other methods can be found in Hayashida \etal (1998) and are
consistent with this general range. 
In \figblr we plot 
the  \fekalfa line width  constraints 
(90\% confidence range) that we measured from the \chandra
data on a diagram of reverberation time-delays versus velocity widths
of some of the BLR optical/UV emission lines
measured by Peterson \& Wandel (1999).
Using their value of the central mass and our 
\fekalfa FWHM of $4515 ^{+3525}_{-2645} \rm \ km/s$
we obtain a size of the  \fekalfa line-emitting region  of 23 light days
($90\%$ confidence, one parameter  limits: $\sim 7-135$ light days).
This places the  \fekalfa line-emitting region predominantly in the outer BLR
and beyond
(from \figblr the BLR extends from $\sim 1-30$ light days).
Since we infer a low ionization
state for the \fekalfa line-emitting
region we would also expect $\rm Mg~{\sc II}~\lambda 2798$ from the same region.
Indeed, Goad \& Koratkar (1998) measure a FWHM of $3581^{+127}_{-161}$
km/s for the broad core of $\rm Mg~{\sc II}~\lambda 2798$.
It is not clear why the innermost region of the BLR is apparently
contributing little to the \fekalfa line. Perhaps the BLR covering factor
increases radially and is significantly less in the inner BLR than
it is in the outer BLR.  
The deduced location of the narrow \fekalfa line emission 
also implies that we might not expect
the line intensity to respond to 
changes in the X-ray continuum intensity on timescales less than a 
week or so. Also, the energy resolution of the HEG does not allow us to rule out
some additional
contribution to the  \fekalfa line from material further out, such as
the putative obscuring torus.

Yaqoob \etal (1993) calculated the expected line profiles,
equivalent widths, and
transfer functions for an  \fekalfa line formed in the BLR.
For pure radial motion in a spherically symmetric
cloud distribution 
(inflow or outflow), the line profile has almost vertical
sides, nearly the same intensity on the red and blue sides
for the narrow widths we are  
considering here. The full width of such a line profile corresponds
to twice the maximum flow velocity.
We fitted the HEG plus MEG data with a very simplistic
rectangular model of the
line profile and derive a flow velocity of $2910_{-1480}^{+4295}$ km/s. 

\subsection{LINE INTENSITY AND EQUIVALENT WIDTH}
\label{eqwidth}

We estimate the intensity and equivalent width of the \fekalfa line
from the BLR following Krolik and Kallman (1987) but use
updated Fe abundance and K-shell cross-section data. 
We assume  a spherically symmetric cloud 
distribution and neutral iron for simplicity.
The line intensity can be obtained by
computing the number of photons 
removed from the continuum above the Fe-K edge threshold
energy (at $E_{K}$, or 7.11 keV for neutral Fe)
and multiplying by the fluorescence yield ($\omega_{K} =0.34$ for
neutral Fe) and the fraction of the sky covered by BLR clouds
($f_{c}$). If the observed power-law continuum is
$N_{p} E^{-\Gamma} \rm \ photons \ cm^{-2} \ s^{-1} $ and
$\sigma_{\rm Fe-K}(E) $ is the K-shell photoelectric absorption
cross-section as a function of energy, then the predicted observed intensity of
the \fekalfa line is

\begin{equation}
I_{\rm Fe-K} = f_{c} \omega_{K} f_{\rm K\alpha} 
\int_{E_{K}}^{\infty} N_{p} E^{-\Gamma} [ 1 - \exp{(-\sigma_{\rm Fe-K}
A_{\rm Fe} N_{H})}]  \  \ \ \ \rm 
photons \ cm^{-2} \ s^{-1}
\end{equation}

where $N_{H}$ is the column density of each cloud, $A_{\rm Fe}$ 
is the Fe abundance relative to Hydrogen, and $f_{\rm K\alpha}$
is the fraction of emission-line photons appearing in the
\fekalfa line
(as opposed to Fe~$K\beta$ line),
and we take $f_{\rm K\alpha}=150/167$ (e.g. Bambynek 1972).
Assuming the clouds are optically thin, we use
a linear approximation for the exponential
in equation 1. We fitted photoelectric absorption cross-section
data from Henke 
tables\footnote{http://www-cxro.lbl.gov/optical\_constants} 
with a simple power-law and
obtained 

\begin{equation}
\sigma_{\rm Fe-K}(E) = 3.79 
\left(\frac{E}{E_{K}}\right)^{-2.646} \times 10^{-20} \ \rm cm^{2}
\ per \ atom. 
\end{equation}

for $E \geq E_{K}$.
This is not as steep as the $(E/E_{K})^{-3}$ relation used in older work
(e.g. Krolik and Kallman 1987; Yaqoob \etal 1993). 
Using an Fe abundance of $4.68 \times 10^{-5}$ (Anders and Grevesse 1989)
the line intensity is then

\begin{eqnarray}
I_{\rm Fe-K\alpha} & =  &
1.04 \times 10^{-5} 
\ \left(\frac{f_{c}}{0.35}\right)
\ \left(\frac{\omega_{K}}{0.34}\right)
\ \left(\frac{\rm A_{\rm Fe}}{4.68 \times 10^{-5}}\right)
\ \left(\frac{N_{H}}{10^{23} \ {\rm cm^{2}}}\right) \times \nonumber \\ 
& & \ \left(\frac{3.2}{\Gamma + 1.646}\right) 
\ \left(\frac{E_{K}}{7.11}\right)^{1-\Gamma}
\ \left(7.11\right)^{1.5-\Gamma}
\ \left(\frac{N_{p}}{4.68 \times 10^{-3}}\right)
\ \rm photons \ cm^{-2} \ s^{-1}.
\end{eqnarray}

The equivalent width of the line is

\begin{eqnarray}
EW_{\rm Fe-K\alpha} & = &
42
\ \left(\frac{f_{c}}{0.35}\right)
\ \left(\frac{\omega_{K}}{0.34}\right)
\ \left(\frac{\rm A_{\rm Fe}}{4.68 \times 10^{-5}}\right)
\ \left(\frac{N_{H}}{10^{23} \ {\rm cm^{2}}}\right)  \times \nonumber \\
& & \ \left(\frac{3.2}{\Gamma + 1.646}\right) 
\ \left(\frac{E_{K}}{7.11}\right)
\ \left(\frac{E_{\rm K\alpha}}{E_{K}}\right)^{\Gamma}
\ \rm eV.
\end{eqnarray}

Here $E_{\rm K\alpha}$ is the center energy of the \fekalfa line.
All quantities in the above equations refer to the source frame. 
The above approximations are valid if the Fe-K
absorption optical depth at all
energies is much less than unity. Effectively this means 
$N_{H} \ll 5.6 \times 10^{23} \rm \ cm^{-2}$ (using the maximum
optical depth, at $E_{K}$). Exact calculation using numerical integration
showed that for $N_{H} = 10^{23} \rm \ cm^{-2}$, the approximations
lead to underestimates of the line intensity and equivalent width of
$\sim 6\%$.

In a recent detailed study involving
modeling the broad optical/UV emission lines in
NGC 5548, Goad and Koratkar (1998) found that
a covering factor of 35\% and a column density of $10^{23} \rm \ cm^{-2}$
satisfy the observational constraints. 
These values, along with
parameters measured for NGC 5548 from the \chandra data,
namely $N_{p} = 4.68 \times 10^{-3}$, $\Gamma = 1.54$, 
give a predicted \fekalfa line intensity 
of $\sim 10^{-5} \rm \ photons \ cm^{-2} \ s^{-1}$ and an EW of
36 eV. These both fall short of the observed values 
($3.6^{+1.7}_{-1.5} \ \rm photons \ cm^{-2} \ s^{-1}$
for the intensity and $133^{+62}_{-54}$ for the EW) 
by a
significant factor.
If the BLR is responsible for the production of the narrow \fekalfa
line, the observed intensity and equivalent width could be reconciled
with the predicted values
if the continuum level was a factor of $\sim 3$ higher in the recent
past before the \chandra observation such that the line had not responded
to the decline in the continuum by the time of the \chandra observation.
Historical values of
the 2--10 keV flux range from $\sim 1.7-6.4 \times \ 10^{-11}
\rm \ ergs \ cm^{-2} \ s^{-1}$ 
(from HEASARC archives and Chiang \etal 2000 and
references therein), so such a scenario is possible.
There is of course the possibility that there is an  
additional contribution to the narrow \fekalfa
line from the putative obscuring 
torus of a few tens of eV.
A torus
component of the line is not
ruled out by our velocity-width measurements and would
bring the observed and predicted strength of the line into
better agreement with a more modest implied change in the continuum level.
Another effect which would increase the predicted equivalent width of
the \fekalfa line is anisotropic X-ray continuum illumination of the
BLR (see Yaqoob \etal 1993 for details).

\section{EFFECT OF THE NARROW LINE ON THE PARAMETERS OF THE
RELATIVISTIC LINE}
\label{ascadata}

NGC 5548, like many other Seyfert~1 galaxies, is known to have
a relativistically broadened  \fekalfa line 
(\fekfig (b); Mushotzky \etal 1995; Nandra \etal 1997). 
The broad line cannot  be clearly seen in 
the \chandra grating data because of the small collecting area and
in \S \ref{broad} we placed formal limits on the intensity of the
broad component from the \chandra data.
We now investigate the effect 
on the relativistic line parameters derived from the \asca data
when a  narrow  \fekalfa emission-line 
component is included.
There have been three observing campaigns of NGC 5548 with \asca
in 1993, 1996 and 1998. The latter two consisted of multiple observations
spaced over one to three weeks and we used the summed data in 
each campaign. We fitted the 3--10 keV spectra with a power law plus
a  model for the  \fekalfa emission line originating in a disk rotating about a
Schwarzschild black hole as
described in \S \ref{diskinaxaf}. No additional
narrow  \fekalfa line was included at first. 
As in \S \ref{diskinaxaf} we fixed the disk-line parameters to
which the data are insensitive, namely the power-law emissivity index at $-2.5$,
the inner and outer disk radii  at $6_{r_{g}}$ and $1000 r_{g}$,
respectively, and the energy of the line in the disk rest-frame at 6.4 keV. 
The inclination angle ($\theta_{\rm obs}$) and the disk-line intensity
($I_{D}$) were allowed to float. Acceptable fits were obtained
for all three spectra and  the results are shown in Table 2.
It can be seen that $\theta_{\rm obs}$ is consistent with a single
value (the weighted mean is $31.5 ^{\circ} \pm 6.5 ^{\circ}$) but the EW during the
July 1998 observation was almost a factor 2 less than that in the
other two epochs, implying time delays between 
variation in the continuum and
line intensity. Indeed the line intensity was less in this observation
than in the previous two observations  which both had a lower continuum
flux. 

Next, we added a narrow Gaussian to the model, with center energy and
width fixed at the values derived from the HEG plus MEG measurements
(Table 1). The
intensities of both the broad and narrow components were allowed
to float. The results are shown in Table 3.
The disk inclination angle is consistent with the spectral fits 
which did not include a narrow line (Table 2) but it is poorly constrained
in the 1993 and 1998 observations. The best constraint, from the
1996 observation is $\theta_{\rm obs} = 31 \pm 8$ degrees.
It can be seen from Table 3 that all the \asca data 
for 1993 and 1998 are consistent with the
existence of a narrow  \fekalfa line  
with the same intensity measured by {\it Chandra}. However, for the 1996
observation, the 90\% confidence, three-parameter
errors on the narrow-line intensity do
not formally overlap with the \chandra measurement. 
We conclude (with caution because of uncertainties
in absolute calibration), that the narrow-line intensity is
variable and was less during the 1996 \asca observation than the
values measured by \chandra and by \asca in 1993 and 1998.
This conclusion is supported by the fact that $\Delta \chi^{2}$ decreases
by 8.1 and 76.9 (for an additional degree of freedom) when
a narrow line is included in the 1993 and 1998 \asca 
dual-line spectral fits but $\Delta \chi^{2} = 0$ for the 1996 \asca data
(see Table 3). 

\section{IMPLICATIONS FOR VARIABILITY OF THE \fekalfa LINE IN SEYFERT~1
GALAXIES}
\label{variability}

The intensity of the disk line derived from the
\asca dual-line spectral
fits (\S \ref{ascadata}) appears to be variable,
it naturally being less for 
larger values of 
the intensity of the best-fitting narrow-line component (Table 3).
Although it appears that both the intensity and EW of the disk line
may be variable amongst the three \asca observations, the 90\%
confidence errors formally overlap so we cannot draw any firm 
conclusions. If the disk-line variability is real then the 
line intensity is not simply correlated with the observed
continuum level because the 1998 observation, the one with the highest 
continuum flux, appears to have the smallest disk-line intensity. 
Time delays between the continuum and line variability
are expected to be less than 10 hours for a black hole mass of
$7 \times 10^{8} M_{\odot}$ 
(appropriate for NGC 5548; e.g. Peterson and Wandel, 1999)
for line emission from within $100r_{g}$.
Since the continuum does not vary significantly on this
timescale during the observing periods this implies that the
response time of the disk line may be longer than expected, implying
more emission from larger radii than is assumed. 
Thus, the broad 
disk-line variability, if it is real,
has difficulty being reconciled with the simple
disk model.
Chiang \etal (2000) actually found that the total Fe-K line intensity
(as measured by {\it RXTE}) is anticorrelated with the 
X-ray continuum flux.
Now we have shown that 
when one takes account of the narrow \fekalfa line the problem is
worse, since 
Table 2 shows that even without a narrow \fekalfa line
the disk line EW in July 1998 is already less than expected
( and any additional
narrow \fekalfa line can only decrease the disk-line intensity).

Since the continuum
level during the \chandra observation was 
at least a factor $\sim 2$ less
than that during the \asca observations 
we might expect the disk line
to have a smaller intensity in the \chandra data. 
However, in \S \ref{broad} we showed that the \chandra data
cannot constrain the broad-line component very well and
obtained an upper limit of $7.6 \times 10^{-5} \rm \ photons \ cm^{-2} \ s^{-1}$
on its intensity, formally consistent with the values obtained
from all the \asca observations.

Recent campaigns to investigate the short-term variability of the 
\fekalfa line in Seyfert~1 galaxies and its relation to the
continuum variability have been conducted for a handful of sources
(NGC 7314, Yaqoob \etal 1996;
MCG $-$6-30-15, Iwasawa \etal 1996, 1999,
Reynolds 2000; NGC 3516, Nandra \etal 1999; NGC 4051, Wang \etal 1999;
NGC 5548, Chiang \etal 2000; NGC 7469, Nandra \etal 2000).
If one assumes that all or most of the \fekalfa line emission comes
from the putative accretion disk then one expects the line intensity to
track the continuum on timescales appropriate for the
size and geometry of
of the X-ray source and disk system. 
Evidence for this is found only in NGC 7314 and NGC 7469.
The other sources show complex behavior which cannot be 
interpreted in terms of this simple model. Even in NGC 7314, only
the broad wings of the \fekalfa line respond to the continuum,
leaving a constant intensity core centered at 6.4 keV (Yaqoob \etal 1996).
This implies that NGC 7314 also has a separate, non-disk narrow
\fekalfa emission line component, just like NGC 5548.
The NGC 7469 data were taken with {\it RXTE} so do not have the
spectral resolution to identify a constant core component of
the emission line.
Moreover, no study has yet shown the expected correlation between the 
\fekalfa line intensity and the Compton-reflection continuum
(which also comes from the accretion disk). This is such a
fundamental prediction of the disk model for the \fekalfa line,
yet when it has been tested either no correlation is found, or
else an {\it anti-correlation} has been found
(such behavior is actually found in NGC 5548; see
Chiang \etal 2000 and \S \ref{ascadata}). 
If a narrow, non-disk \fekalfa line component  becomes important in general
for low-flux states of Seyfert~1 galaxies, then variability studies 
need to take this into account.
While discussions of a non-disk component
to the \fekalfa lines have been around ever since their discovery 
in Seyfert~1 galaxies
with early proportional counters in the late seventies, the \chandra 
observation of NGC 5548 now provides direct observational proof that
this component exists, at least in one source.

\section{CONCLUSIONS}
\label{concl}

Using the \chandra High-Energy Grating we detected a narrow
 \fekalfa emission line in the Seyfert~1 galaxy NGC 5548,
with a center energy of $6.40 \pm 0.03$ keV,
a FWHM of $4515^{+3525}_{-2645}$ km/s and an equivalent
width (EW) of $133 ^{+62}_{-54}$ eV 
(90\% confidence errors, three parameters of
interest; all quantities are in the galaxy frame). The
line is only marginally resolved.
This emission line is distinct from the broad \fekalfa line 
in NGC 5548 and other Seyfert~1 galaxies,
thought to originate in a
disk rotating around a central black hole. The \chandra data
are inconsistent with the line originating only from a disk
because it is too sharply peaked even for low-inclination disks.  
The narrow \fekalfa line 
center energy, velocity width and equivalent width
measurements are consistent
with the bulk of the
line originating in cold, neutral matter in the outer BLR
at about a light-week to a few light-months from the central source.
We also do not, however, rule out ionization states up to about Fe~XVIII.

The agreement of the line equivalent width and intensity with 
predicted values 
assumes a spherically symmetric distribution of
BLR clouds with a covering factor of 35\% and typical cloud
column densities of $\sim 10^{23} \ \rm cm^{-2}$ 
(values consistent with optical/UV emission-line studies) and
is conditional upon the X-ray continuum  
having been a factor $\sim 3$ higher in the immediate past prior
to the \chandra observation. The 2--10 keV flux during the
\chandra observation was a factor of 2.5 below typical
historical values so this is not
unreasonable. The continuum variability constraint would be
relieved somewhat if there were a 
contribution to the
narrow line from a putative,
parsec-scale obscuring torus, which has been postulated as a key
component of unification schemes of Seyfert~1 and Seyfert~2 galaxies.
Indeed, such a contribution is not ruled out by the data.
Anisotropic X-ray illumination of a spherical BLR would also
increase the expected equivalent width of the narrow line.
Considering its deduced location,
the narrow  \fekalfa line iteself may vary
on timescales of weeks to months. Indeed, examination of
archival \asca data provides evidence for variability of the
narrow \fekalfa line.

We do not yet know 
whether  the narrow  \fekalfa emission line is a common feature of
Seyfert~1 galaxies. From a moderate sample of sources, 
Nandra \etal (1997) placed an upper limit on a narrow  \fekalfa component
in the overall  \fekalfa line profile typically of $\sim 60$ eV, but
some sources show stronger
evidence for a separate, narrow  \fekalfa line
(e.g. Yaqoob \& Weaver 1996; Weaver and Reynolds 1998; Done \etal 2000).
In the case of NGC 5548 one of
the likely reasons for the relatively large
EW of the narrow component observed
by \chandra 
is that the continuum intensity happened to be  
significantly lower during the \chandra observation
compared to the upper range of historical values. 
It will be important to obtain \chandra data for
a large sample of Seyfert~1 galaxies.
However, studying the variability of the narrow line is rather expensive with
\chandra.
Monitoring the narrow  \fekalfa line
and using it for reverberation mapping 
with future
X-ray astronomy missions with larger effective area will allow
us to independently measure the central black-hole mass, complementing 
reverberation studies
using optical/UV emission lines. 
Combining these future variability studies with spectral analysis with
an 
energy of resolution of the order
of 1 eV will allow us to 
constrain the details of the geometry and dynamics of the
gas which are not accessible by current observations.

We thank Dr. Kimberly Weaver and Professor Julian Krolik for 
valuable discussions relating to this work.
We also thank Drs. H. Marshall and F. Nicastro for their helpful advice on
\chandra data analysis issues. We are grateful to an anonymous
referee for reviewing the manuscript and making some
important suggestions. This research made use of HEASARC online 
databases.   

\newpage

\begin{deluxetable}{lcc}
\tablecaption{ Parameters of the narrow  \fekalfa Line Derived from  the HEG Data}
\tablecolumns{3}
\tablewidth{0pt}
\tablehead{
\colhead{Parameter} & \colhead{HEG only } & \colhead{HEG and MEG}} 

\startdata

Center Energy (keV) & $6.390 ^{+0.032}_{-0.029}$ &
			$6.402 ^{+0.027}_{-0.025}$ \nl

Gaussian width (eV) & $41 ^{+37}_{-29}$  & 
			$41 ^{+32}_{-24}$  \nl

Intensity 
($10^{-5} \ \rm photons \ cm^{-2} \ s^{-1}$)
& $3.2 ^{+1.8}_{-1.6} $ & $3.6 ^{+1.7}_{-1.5} $ \nl

Equivalent Width (EW) & $120^{+40}_{-38}$ &
			$133^{+62}_{-54}$  \nl

Velocity FWHM (km/s) & $4525^{+4080}_{-3200}$ & $4525^{+3525}_{-2645}$ \nl

Photon Index ($\Gamma$) & $1.531 ^{+0.087}_{-0.088}$ &
			$1.540 ^{+0.060}_{-0.059}$ \nl

C-statistic & 842.7 & 1332.8 \nl

degrees of freedom & 1765 & 2648 \nl

2--10 keV flux
($10^{-11} \ \rm ergs \  cm^{-2} \ s^{-1}$) & 2.5 & 2.4 \nl 

\tablecomments{All line parameter values are referred to the rest frame
of NGC 5548 ($z=0.01676$). Statistical errors are
90\% confidence for one interesting parameter
($\Delta C = 2.706$). The 2--10 keV flux for the HEG plus MEG 
fit is the average from the two gratings.}

\enddata
\end{deluxetable}

\newpage

\begin{deluxetable}{lccc}
\tablecaption{ Relativistic Disk Line Fits to \asca Data for \psrc}
\tablecolumns{4}
\tablewidth{0pt}
\tablehead{
\colhead{Parameter} & \colhead{July 1993} &
                \colhead{July 1996} & \colhead{July 1998}
}

\startdata

Exposure (ks) & 29.6 & 56.9 & 130.0 \nl

Disk inclination/ $\theta_{\rm obs}$ (degrees)
& $33^{+14}_{-12}$ & $31^{+7}_{-8}$ 
		 & $29^{+9}_{-12}$ \nl

Disk-line intensity ($\rm 10^{-5} \ photons \ cm^{-2} \ s^{-1}$) 
 & $10.3^{+3.7}_{-3.6}$    & 
		$11.3^{+3.3}_{-3.2}$ 
	  &   $7.6^{+2.2}_{-2.1}$ \nl               
Disk-line EW (eV) & $234^{+84}_{-81}$ 
        &                  $208^{+60}_{-58}$ 
        & $120_{-36}^{+34}$ \nl

2--10 keV flux ($10^{-11} \rm \ ergs \ cm^{-2} \ s^{-1}$)
 & 5.1 & 5.4 & 6.5 \nl

$\chi^{2}$ & 1016.3 & 1199.8 &  1483.9 \nl

degrees of freedom & 1004 & 1195 & 1466 \nl

& & \nl

\enddata

\tablecomments{\small
3--10 keV spectra fits using \asca data
(both SIS and GIS) for three observations of NGC 5548.
The model consists of a power law plus an emission line
(rest energy 6.4 keV)
from a relativistic disk rotating around a Schwarzschild black hole. 
The inner and outer disk radii are
fixed at 6 and 1000 gravitational radii respectively. The
line emissivity is a power law with index fixed at $-2.5$.
All the line parameters shown refer to the galaxy frame. 
All statistical errors on parameters 
are 90\% confidence for three interesting parameters
($\Delta \chi^{2} = 6.251$).}
\end{deluxetable}

\newpage

\begin{deluxetable}{lccc}
\tablecaption{ Dual Emission-Line Model:
Relativistic Disk Line Plus Gaussian Fits to \asca Data for \psrc}
\tablecolumns{4}
\tablewidth{0pt}
\tablehead{
\colhead{Parameter} & \colhead{July 1993} &
                \colhead{July 1996} & \colhead{July 1998}
}

\startdata

Disk inclination/ $\theta_{\rm obs}$ (degrees)
& $46^{+14}_{-28}$ & $31^{+8}_{-8}$ 
		 & $52^{+38}_{-52}$ \nl

Disk-line Intensity ($\rm 10^{-5} \ photons \ cm^{-2} \ s^{-1}$) 
 & $7.1 ^{+5.5}_{-5.6}$    & 
		$11.3^{+3.2}_{-4.0}$ 
	  &   $5.5^{+4.1}_{-4.2}$ \nl               
Disk-line EW (eV) & $147^{+114}_{-116}$ 
        &                  $208^{+53}_{-73}$ 
        & $82_{-63}^{+63}$ \nl

Narrow-Line  Intensity ($\rm 10^{-5} \ photons \ cm^{-2} \ s^{-1}$)
 & $3.2^{+2.2}_{-2.8}$ & $0.0^{+1.8}_{-0.0}$ & $2.9^{+1.4}_{-1.5}$ \nl

Narrow-Line (eV) & $63^{+43}_{-55}$ & $0^{+27}_{-0}$ 
		& $41^{+20}_{-20}$ \nl

$\chi^{2}$ & 1008.2 & 1199.8 &  1407.7 \nl

degrees of freedom & 1003 & 1194 & 1428 \nl

\enddata
\tablecomments{\small
Spectral fits to \asca data for NGC 5548 using
a composite model for the \fekalfa line,
consisting of a relativistic disk line plus a narrow Gaussian.
The Gaussian width of the narrow line is fixed at the
	best-fitting value obtained from the HEG plus MEG
\chandra spectral fits ($\sigma = 41$ eV, Table 1).
All line parameters shown refer to the galaxy frame.
All statistical errors  
are 90\% confidence for three interesting parameters
($\Delta \chi^{2} = 6.251$).
}
\end{deluxetable}

\newpage
\section*{Figure Captions}

\par\noindent
{\bf Figure 1} \\
(a) \chandra \hetgs spectra for NGC 5548 from the HEG (solid error
bars) and
MEG (crosses for data points with dotted error bars).
Solid line is the best-fitting power-law
continuum to HEG plus MEG data but with the best-fitting HEG
normalization. The
$\pm 1$ orders are summed.
The narrow  \fekalfa
emission-line component is clearly detected in both the HEG and MEG.
(b) \asca SIS0+SIS1 relativistic  \fekalfa line profile
for the July 1996 observation of
NGC 5548 (see text). (c)
Close-up of the HEG spectrum; the data
are smoothed with a boxcar  five bins wide, where the binsize is
$0.005 \AA$.
The solid line is the best-fitting model which
consists of a power-law continuum and a narrow Gaussian.
All 
parameters have the best-fitting values derived from
fitting the HEG data (Table 1).
Energy scale in all three panels
is for the {\it observer's} frame.

\par\noindent
{\bf Figure 2} \\
Joint two-parameter 
confidence region of the narrow \fekalfa line intensity versus
Gaussian width from spectral fitting to the HEG and MEG \chandra
data. The contours correspond to
68\%, 90\% and 99\% confidence.

\par\noindent
{\bf Figure 3} \\
Measurements by Peterson and Wandel (1999) of the time lags (relative
to the appropriate continuum ) and
the FWHM of various optical/UV emission lines in NGC 5548. The dashed 
line corresponds to the best Keplerian model fit 
obtained by Wandel and Peterson (1999) for the data from which they
obtain a central mass of $6.8 \times 10^{7} \ M_{\odot}$.
The dotted lines show the best-fitting
parameters and 90\% confidence (one-parameter) limits on the
FWHM of the narrow \fekalfa emission line 
measured from our \chandra HEG plus MEG data (see Table 1). 

\begin{figure}[h]
\vspace{10pt}
\centerline{\psfig{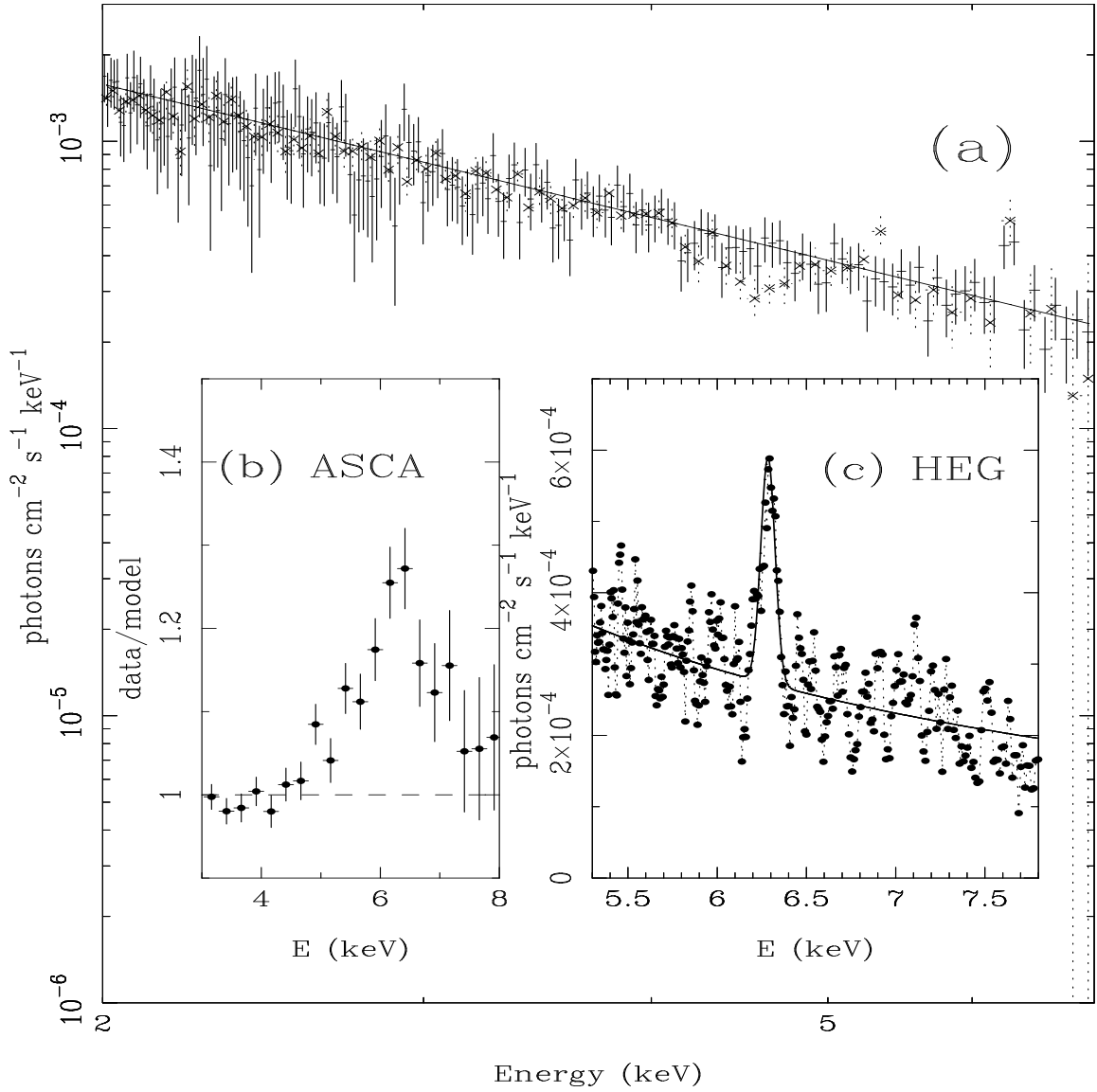}
}
\caption{See figure caption. }
\end{figure}

\begin{figure}[h]
\vspace{10pt}
\centerline{\psfig{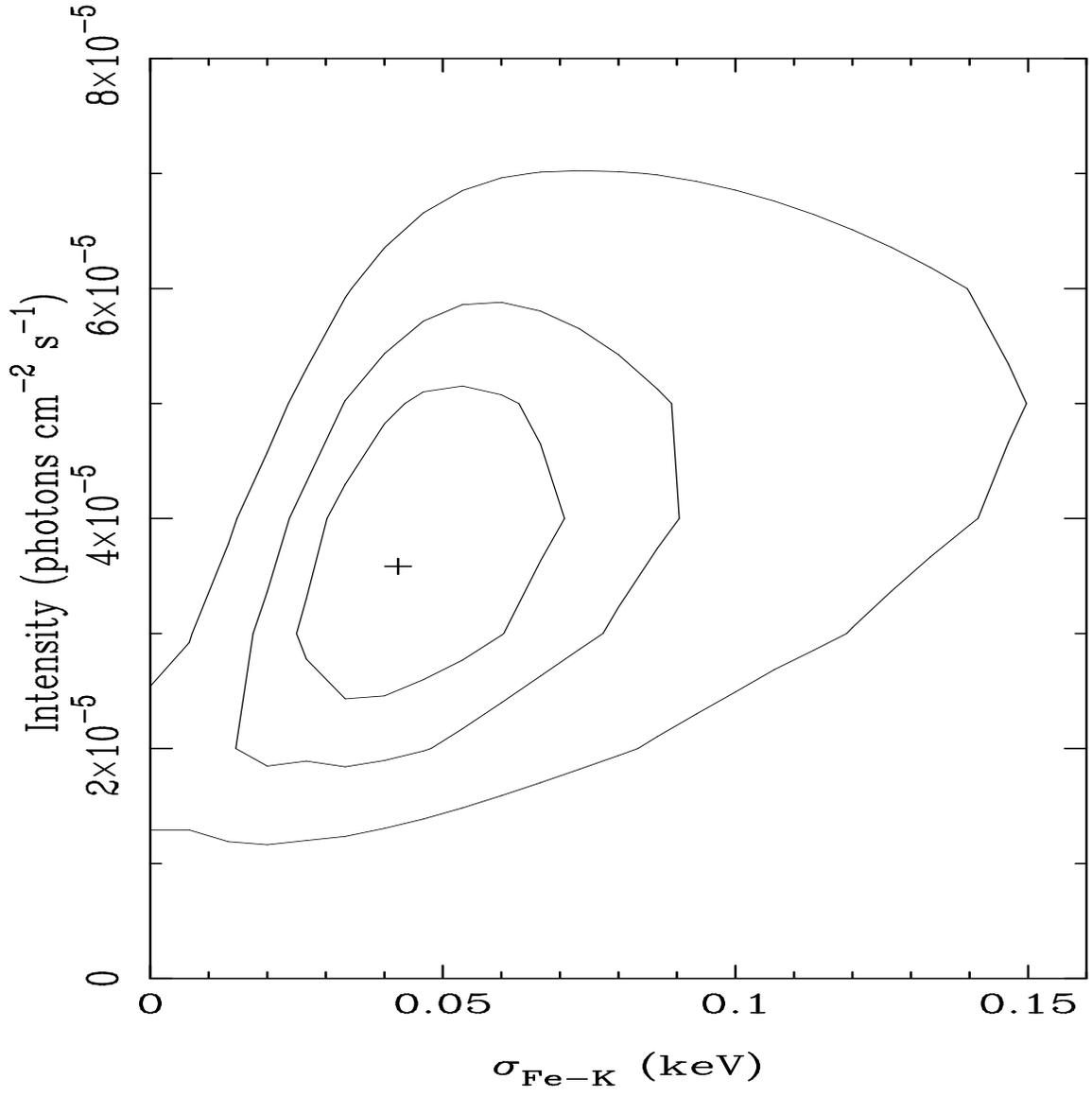}
}
\caption{See figure caption.}
\end{figure}

\begin{figure}[h]
\vspace{10pt}
\centerline{\psfig{file=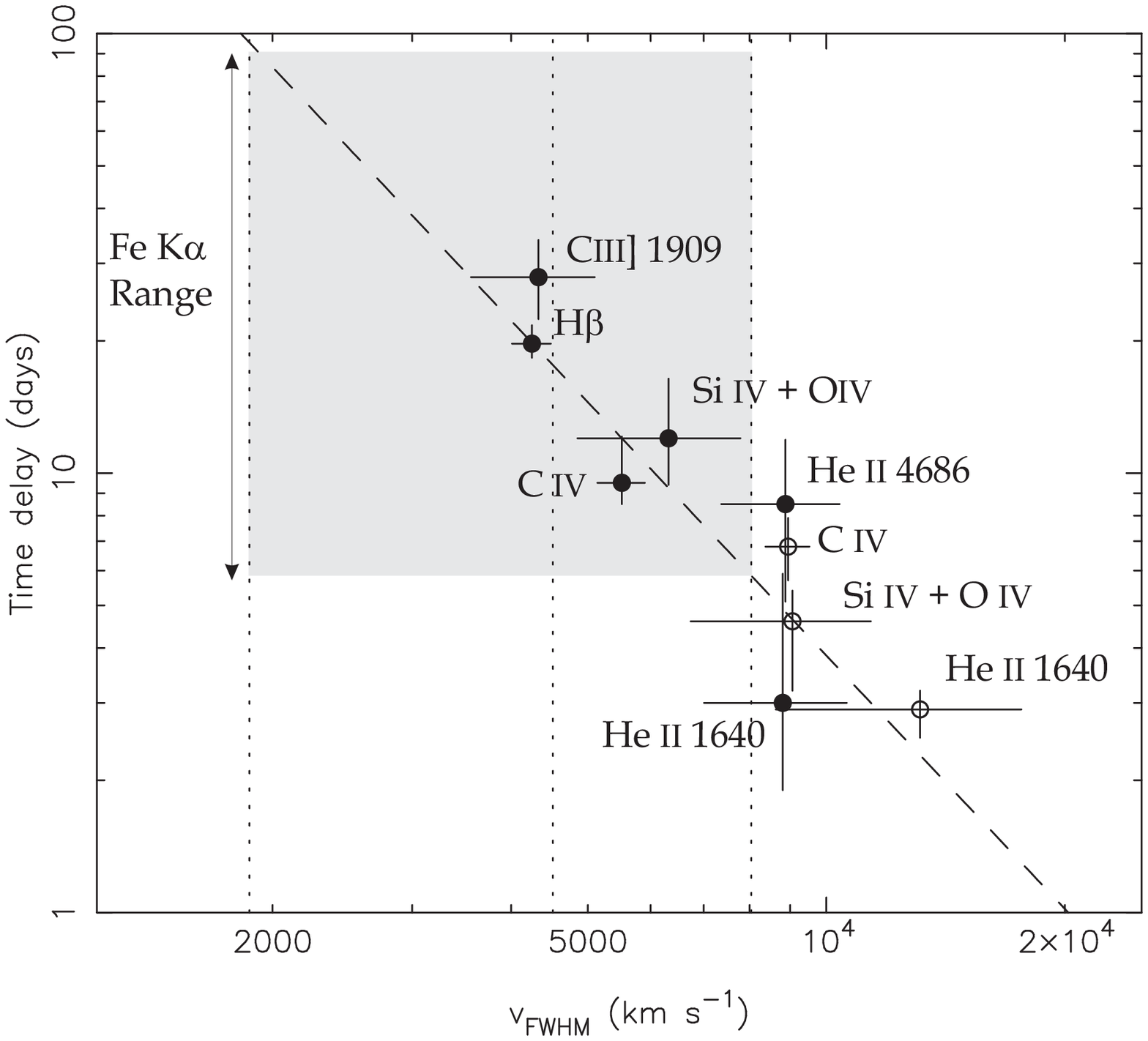,width=6.0in,height=6.0in}
}
\caption{See figure caption. }
\end{figure}

\end{document}